\documentclass[aps,prb,reprint,superscriptaddress]{revtex4-2}

\usepackage{graphicx}%
\usepackage{amsmath,amssymb,amsfonts}%
\usepackage{xcolor}%
\usepackage{siunitx}
\usepackage[T1]{fontenc} 
\usepackage{upgreek}
\usepackage[colorlinks=true,linkcolor=blue,anchorcolor=blue, citecolor=blue,urlcolor=blue]{hyperref}%

\newcommand{\addressA}{National Laboratory of Solid State Microstructures, Collaborative Innovation Center of Advanced Microstructures, School of Physics, Nanjing University, Nanjing 210093, China}
\newcommand{\addressB}{Center for Quantum Matter, School of Physics, Zhejiang University, Hangzhou 310058, China}
\newcommand{\addressC}{Research Center for Electronic and Optical Materials, National Institute for Materials Science, 1-1 Namiki, Tsukuba 305-0044, Japan}
\newcommand{\addressD}{Research Center for Materials Nanoarchitectonics, National Institute for Materials Science, 1-1 Namiki, Tsukuba 305-0044, Japan}
\newcommand{\addressE}{Jiangsu Physical Science Research Center, Nanjing University, Nanjing 210093, China}

\begin{document}
	\title{Tailoring pure valley-Zeeman spin-orbit coupling in WSe$_2$-encapsulated monolayer graphene}
	
	\author{Yaqing Han}
	\altaffiliation{These authors contributed equally to this work.}
	\author{Siqi Jiang}
	\altaffiliation{These authors contributed equally to this work.}
	\author{Jingkuan Xiao}
	\altaffiliation{These authors contributed equally to this work.}
	\affiliation{\addressA}
	
	\author{Jiawei Jiang}
	\affiliation{\addressA}
	\affiliation{\addressB}
	
	\author{Yulu Liu}
	\author{Jiabei Huang}
	\author{Yu Du}
	\author{Di Zhang}
	\author{Fuzhuo Lian}
	\author{Wanting Xu}
	\author{Siqin Wang}
	\affiliation{\addressA}
		
	\author{Kenji Watanabe}
	\affiliation{\addressC}
	
	\author{Takashi Taniguchi}
	\affiliation{\addressD}
	\author{Xiaoxiang Xi}
	\author{Alexander S. Mayorov}
	\author{Renjun Du}
	\altaffiliation{Contact author:~renjundu@nju.edu.cn}
	\affiliation{\addressA}
	
	\author{Kai Chang}
	\author{Hongxin Yang}
	\affiliation{\addressB}
	
	\author{Lei Wang}
	\altaffiliation{Contact author:~leiwang@nju.edu.cn}
	\author{Geliang Yu}
	\altaffiliation{Contact author:~yugeliang@nju.edu.cn}
	\affiliation{\addressA}
	\affiliation{\addressE}
	
	\begin{abstract}
		Engineering proximity effects in twisted van der Waals heterostructures offers a powerful platform for designing electronic properties.
		While theoretical predictions of quantum interference in transition metal dichalcogenide-encapsulated graphene can selectively control the spin-orbit coupling component, experimental realizations have remained elusive.
		Here, we report pure valley-Zeeman spin-orbit coupling in monolayer graphene, achieved by encapsulation between two parallel twisted WSe$_2$ monolayers.
		We observed a symmetry-enforced reordering of Landau levels, which is driven by the competition between the fixed valley-Zeeman energy and the magnetic-field-dependent cyclotron energy.
		This reordering is characterized by a transition from symmetry-broken states in the quantum Hall effect to a restored fourfold degeneracy with integer or half-integer quantum Hall sequences.
		We also demonstrate the ability to completely quench the proximity spin-orbit coupling by tuning the encapsulated geometry.
	\end{abstract}
	
	\maketitle
	
	Proximity-induced modification of electronic properties in van der Waals (vdW) heterostructures has emerged as a powerful platform for engineering quantum materials with tailored functionality~\cite{geim2013Van, andrei2020Graphene, kennes2021Moire, li2023Proximityinduced, pandey2025ProximityMediated, zollner2025Firstprinciples, 0p002}.
	In particular, imprinting strong spin-orbit coupling (SOC) into graphene via adjacent transition metal dichalcogenides (TMDC)~\cite{avsar2014Spin,  han2014Graphene, wang2015Strong,kim2015Band, wang2016Origin,benitez2018Strongly, hu2024Tunable,han2024Large,zhang2025Twistprogrammable} has opened new pathways to explore exotic phenomena such as spin transports~\cite{benitez2018Strongly}, ferromagnetism~\cite{han2024Large}, and unconventional superconductivity~\cite{zhang2025Twistprogrammable}.
	This proximity effect typically induces a mixture of SOC components with different symmetries, such as valley-Zeeman and Rashba terms~\cite{tiwari2022Experimental, rao2023Ballistic, masseroni2024Spinorbit, yang2024Twistangletunable}.
	However, disentangling these contributions and selectively controlling individual SOC components remain elusive, yet are essential for realizing advanced spintronic and topological devices.
	
	The twist-angle between adjacent layers provides a powerful knob for tuning the proximity effects~\cite{david2019Induced, li2019Twistangle, lee2022Chargetospin, naimer2023Twistangle, yang2024Twistangletunable, naimer2024Tuning}.
	In the graphene/TMDC bilayer, twisting modulates the strength of SOC and introduces a spin texture~\cite{li2019Twistangle}.
	For instance, a misalignment of 30$^\circ$ is expected to restore the mirror plane symmetry that quenches the valley-Zeeman component~\cite{lee2022Chargetospin}.
	Recent theoretical works~\cite{island2019Spin, peterfalvi2022Quantum, zollner2023Twist} predict that TMDC-encapsulated graphene enables quantum interference between the top and bottom proximity SOC, offering a route to selectively amplify or suppress a specific SOC term.

	In this Letter, we engineered pure valley-Zeeman SOC in monolayer graphene (MLG) by symmetrically encapsulating MLG between two WSe$_2$ layers.
	Through magneto-transport measurements, we observe the symmetry-enforced reordering of Landau level (LL) spectra from symmetry-broken quantum Hall states to a restored fourfold degeneracy, characterized by half-integer or integer quantization sequences.
	Our tight-binding model precisely reproduces LL spectra and establishes it as a characteristic for engineering SOC.
	We further show that the proximity SOC can be completely suppressed by tuning the encapsulated geometry.
	Our results provide experimental evidence of the tailored SOC in WSe$_2$-encapsulated MLG, opening avenues for exploring topological and spintronic phases in vdW heterostructures.
	
	\begin{figure*}[htbp]
		\includegraphics[scale=1]{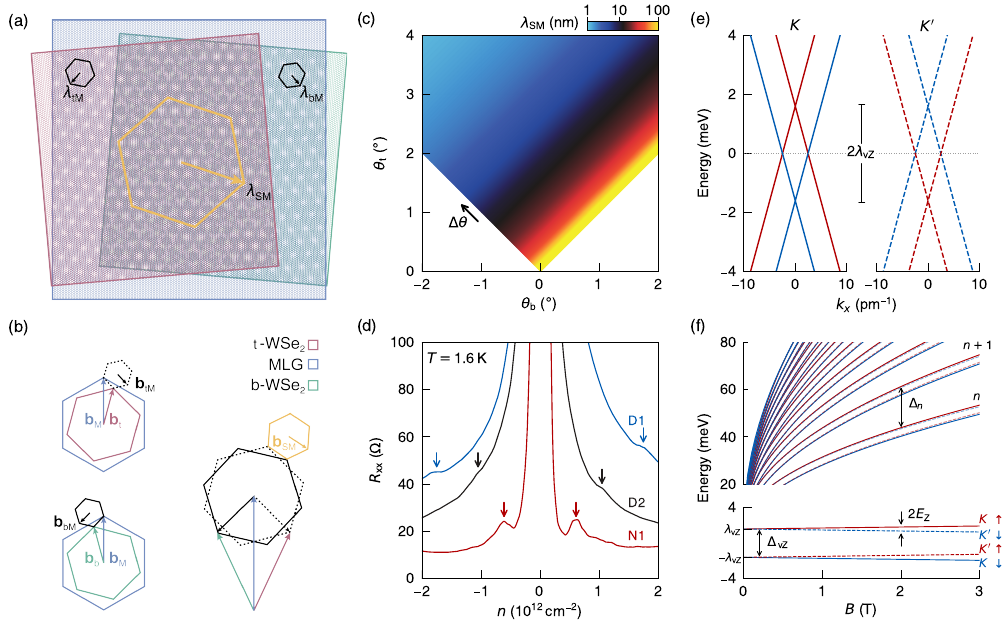}
		\caption{(a) Schematic of the WSe$_2$/MLG/WSe$_2$ super-moir\'e lattice. 
		Top- and bottom-WSe$_2$ monolayers are twisted relative to MLG by $\theta_\mathrm{t}$ and $\theta_\mathrm{b}$, constructing moir\'e superlattices with wavelengths of $\lambda_\mathrm{tM}$ and $\lambda_\mathrm{bM}$, respectively. Their overlap forms a super-moir\'e lattice with $\lambda_\mathrm{SM}$.
		(b) Schematic of Brillouin zones of top-WSe$_2$ (red), MLG (blue) and bottom-WSe$_2$ (green). 
		The reciprocal vectors of the primary moir\'e (super-moir\'e) lattices are denoted by black (yellow) arrows.
		(c) Calculated $\lambda_\mathrm{SM}$ as functions of $\theta_\mathrm{b}$ and $\theta_\mathrm{t}$ for a small twist-angle between WSe$_2$ layers $\Delta \theta = \left| \theta_\mathrm{t} - \theta_\mathrm{b} \right|$.
		(d) $R_\mathrm{xx}$ as a function of $n$ in investigated devices measured at $B=0\,\mathrm{T}$ and $T=1.6\,\mathrm{K}$.
		Arrows indicate the positions of SDPs.
		(e) Band structures of MLG near the $K$ and $K^\prime$ points of the Brillouin zone with proximitized $\lambda_\mathrm{vZ}=1.6\,\mathrm{meV}$.
		(f) Calculated LL spectrum for $B>0$ in MLG with pure valley-Zeeman SOC, consisting of out-of-plane spin-up (red) and spin-down (blue) LLs.
		The solid and dashed lines denote LLs from $K$ and $K^\prime$ valleys, respectively.
		The two spin-valley-locked sub-bands are separated by $\Delta_{\mathrm{vZ}}=2\lambda_{\mathrm{vZ}}$.
		$\Delta_n$ indicates the spacing between adjacent LLs for a given valley and spin flavor, and $E_\mathrm{Z}=\frac{1}{2}g\mu_B B$ denotes the $B$-induced conventional Zeeman energy.
		}
		\label{fig:Fig1}
	\end{figure*}
	
	We fabricated vdW heterostructures featuring MLG encapsulated by two aligned WSe$_2$ monolayers, as depicted in Fig.~\ref{fig:Fig1}(a).
	The constituent WSe$_2$ flakes are derived from the same parent monolayer and assembled by a 'tear-and-stack' transfer method~\cite{kim2016Van} to ensure that the twist-angle $\Delta \theta$ between the WSe$_2$ layers is small, which can be determined from their distinct second harmonic generation (SHG) polarization patterns (see Supplemental Material~\uppercase\expandafter{\romannumeral5}~\cite{See}\nocite{zubair2020Influence, kumar2021Zerofield, sichau2019Resonance, yankowitz2012Emergence, li2013Probing, rickhaus2019Gap, taychatanapat2013Electrically}).
	Due to the large lattice mismatch, the individual WSe$_2$/MLG interface exhibits a negligible moir\'e wavelength ($\lambda < 1 \,\mathrm{nm}$).
	However, for WSe$_2$/MLG/WSe$_2$, the coherent electronic coupling across MLG enables interference of the two moir\'e patterns [Fig.~\ref{fig:Fig1}(b)], giving rise to a long-wavelength potential modulation known as a super-moir\'e lattice~\cite{hermann2012Periodic, wang2019Newa, jiang2025Interplay}.
	The calculated $\lambda_\mathrm{SM}$ shown in Fig.~\ref{fig:Fig1}(c), plotted as functions of the twist-angles of top-WSe$_2$/MLG ($\theta_\mathrm{t}$) and MLG/bottom-WSe$_2$ ($\theta_\mathrm{b}$), reveal an inverse dependence on $\Delta \theta = \left| \theta_\mathrm{t} - \theta_\mathrm{b} \right|$ (calculation detailed in Supplemental Material~\uppercase\expandafter{\romannumeral4}~\cite{See}).
	
	The electronic signature of this sandwiched super-moir\'e structure is observed in our low-temperature transport measurements. 
	Here, we investigate three cross-layer-aligned devices with distinct interlayer alignment.
	As indicated by arrows in Fig.~\ref{fig:Fig1}(d), the longitudinal resistance traces $R_\mathrm{xx}$ exhibit satellite peaks on both sides of the charge neutrality point (CNP), located symmetrically at carrier densities $\pm n_\mathrm{SM}$ and signaling the presence of secondary Dirac points (SDPs) induced by the super-moiré potential.
	According to the positions of SDPs, we can extract super-moir\'e wavelengths ranging from 15 to $28\,\mathrm{nm}$ for our devices, in well agreement with our SHG and crystallographic analysis. 	
	
	Crucially, the rotational alignment of WSe$_2$ monolayers with respect to the graphene lattice acts as a control parameter for the electronic properties. 
	This global alignment determines the interlayer coupling momenta within the WSe$_2$ Brillouin zone, governing the atomic and orbital hybridization at interfaces~\cite{li2019Twistangle} and controlling the proximity SOC in MLG~\cite{yang2024Twistangletunable}.
	Theoretically, this coupling is primarily described by two terms: the valley-Zeeman interaction, $H_\mathrm{vZ} = \lambda_\mathrm{vZ} \tau_{z} \sigma_0 s_{z}$, and the Rashba term, $H_\mathrm{R} = \lambda_\mathrm{R} (-\tau_{z} \sigma_x s_{y} - \sigma_y s_{x})$, where $\tau_{z}=\pm1$ is the valley index, $\sigma_i$ and $s_i$ are Pauli matrices for the sublattice and spin degrees of freedom, respectively.
	The coupling strengths $\lambda_\mathrm{vZ}$ and $\lambda_\mathrm{R}$ are directly modulated by the engineered twist-angle between MLG and WSe$_2$~\cite{david2019Induced}. 
	For all investigated devices, the super-moiré features confirm that the two encapsulating WSe$_2$ layers are nearly aligned.
	In this configuration, the inversion symmetry broken at each interface is effectively restored globally, which generates the opposite phase of the Rashba term~\cite{island2019Spin, peterfalvi2022Quantum, zollner2023Twist}.
	Consequently, the Rashba SOC components from the two interfaces interfere destructively, $\lambda_\mathrm{R} \approx 0$.
	However, the valley-Zeeman term, arising from sublattice symmetry-breaking that is even under inversion, interferes constructively~\cite{zollner2023Twist}.
	The low-energy electronic states can be described by the Dirac Hamiltonian of MLG perturbed by valley-Zeeman SOC, which induces opposite spin splittings in the $K$ and $K^\prime$ valleys while preserving time-reversal symmetry and locking the spin out of plane (see Supplemental Material~\uppercase\expandafter{\romannumeral1}~\cite{See}).
	The corresponding band structure and LL spectrum are shown in Figs.~\ref{fig:Fig1}(e) and~\ref{fig:Fig1}(f).
	This selective amplification of one spin-orbit term while suppressing another demonstrates a remarkable tunability of proximity SOC.

	\begin{figure}[t]
		\centering
		\includegraphics[scale=1]{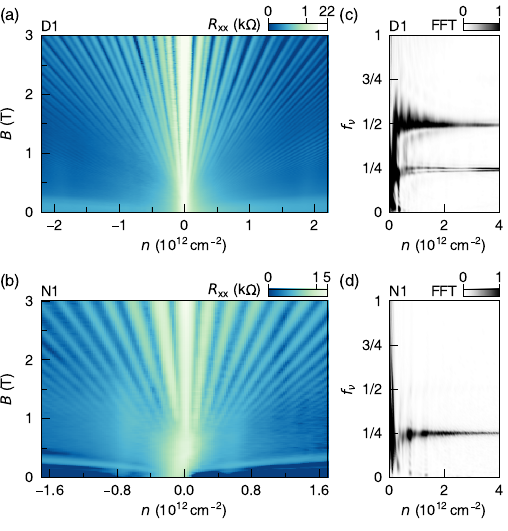}
		\caption{(a),(b) $R_\mathrm{xx}$ as functions of $n$ and $B$ measured at $D=0$ for the devices D1 and N1, respectively.
		(c),(d) FFT spectra obtained from the Landau fans shown in (a) and (b), respectively, plotted as functions of carrier density $n$ and normalized frequency $f_\nu=f/(n\phi_0)$.
		Here, $f$ is the oscillation frequency calculated for each density in Tesla, and $\phi_0=h/e$ is the magnetic flux quantum.}
		\label{fig:Fig2}
	\end{figure}
	
	The Shubnikov-de Haas (SdH) effect originates from the Landau quantization of cyclotron orbits under a perpendicular magnetic field ($B$), and provides a sensitive probe of the Fermi-surface evolution and the underlying LL spectrum.
	By analyzing the quantum oscillations in $R_\mathrm{xx}$ as a function of $B$, we can analyze the engineered SOC in our devices.
	Here, we present a comparative study of two representative devices, D1 and N1, which have similar small $\Delta \theta$ but different global alignments with respect to the graphene lattice.
	As demonstrated in Fig.~\ref{fig:Fig2}, the exceptional quality of our heterostructures is evidenced by the onset of SdH oscillations below $0.3\,\mathrm{T}$. 
	Notably, the symmetry-broken quantum Hall states in device D1 [Fig.~\ref{fig:Fig2}(a)] emerge at $B$ as low as $0.6\,\mathrm{T}$, consistent with the theoretically predicted enhanced valley-Zeeman SOC in this aligned configuration~\cite{peterfalvi2022Quantum} ($\theta_\mathrm{t} \approx 5.4^\circ$ and $\theta_\mathrm{b} \approx 6.6^\circ$).
	To gain insight into this band splitting, we performed fast Fourier transforms (FFT) on magnetoresistance.
	The density-dependent frequency of SdH oscillations $f$ sampled in $1/B$ is normalized as $f_\nu=f/(n\phi_0)$ [Figs.~\ref{fig:Fig2}(c) and \ref{fig:Fig2}(d)], corresponding to the fraction of Luttinger volume enclosed by the Fermi surface~\cite{zhang2025Twistprogrammable}, where $\phi_0=h/e$ is the magnetic flux quantum, $h$ is the Planck constant, and $e$ is the elementary charge.
	For device D1, the FFT spectrum, plotted as functions of the carrier density $n$ and normalized frequency $f_\nu$ in Fig.~\ref{fig:Fig2}(c), reveals two distinct frequencies split from $f_\nu=1/4$, $f_\nu^{+}$ and $f_\nu^{-}$.
	This frequency splitting originates from two Fermi pockets with imbalanced areas~\cite{rao2023Ballistic, masseroni2024Spinorbit}.
	By relating $\Delta f_\nu = \left| f_\nu^{+} - f_\nu^{-} \right|$ to the band splitting, we extracted the effective SOC strength $\lambda_\mathrm{SOC} = \sqrt{\lambda_\mathrm{vZ}^2 + \lambda_\mathrm{R}^2} = 1.68 \pm 0.12\,\mathrm{meV}$ (see Supplemental Material~\uppercase\expandafter{\romannumeral3}~\cite{See}).
	Furthermore, we conducted transverse magnetic focusing (TMF) measurements, in which the persistent splitting of higher-order focusing peaks directly probes the spin-orbit split Fermi surface and confirms the establishment of pure valley-Zeeman SOC regime~\cite{rao2023Ballistic, kang2024Magnetotransport} (details described in Supplemental Material~\uppercase\expandafter{\romannumeral8}~\cite{See})

	\begin{figure*}[htbp]
		\centering
		\includegraphics[scale=1]{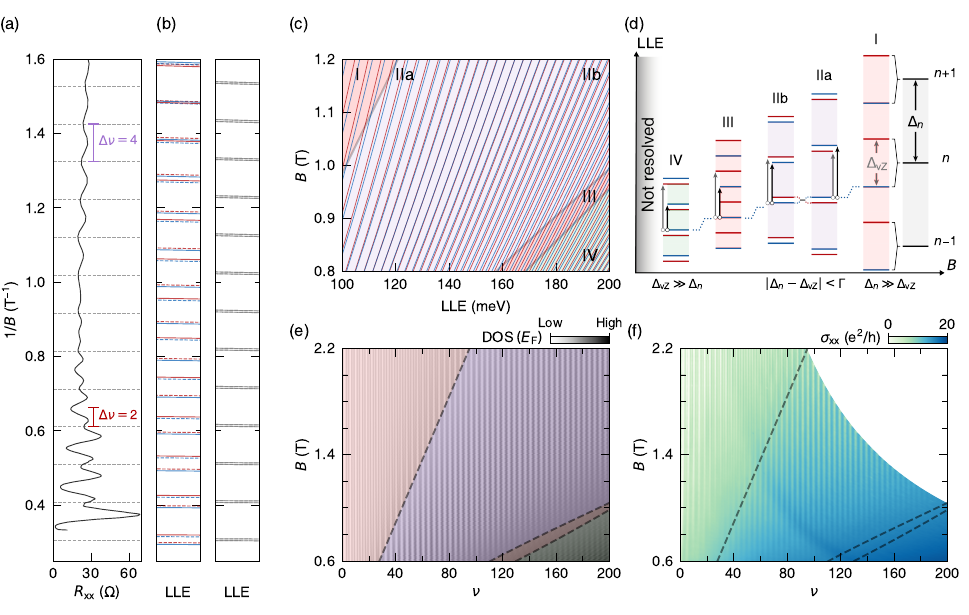}
		\caption{(a) $R_\mathrm{xx}$ as a function of $1/B$ at $n=1.05 \times 10^{12}\,\mathrm{cm}^{-2}$, replotted from Fig.~\ref{fig:Fig2}(a).
		The gray-dashed lines denote the integer quantum Hall sequence of LL filling factor $\nu=4N$, where the index $N$ is an integer.
		(b) Calculated LL energy (LLE) as a function of $1/B$ at $\lambda_\mathrm{vZ}=1.6$ (left-panel) and $0\,\mathrm{meV}$ (right-panel), respectively.
		The color code (red and blue) represents the out-of-plane spin (up and down), and gray denotes the states of pristine MLG.
		Solid and dashed lines denote LLs from $K$ and $K^\prime$ valleys, respectively.
		(c) Calculated LLE as a function of $B$ at $\lambda_\mathrm{vZ}=1.6\,\mathrm{meV}$.
		(d) Schematic of LLs as a function of $B$, depicting the competition between $\Delta_\mathrm{vZ}$ and $\Delta_n$.
		Only the LLs of the $K$ valley are presented in (c,d) to avoid clutter.
		(e) Calculated $\mathrm{DOS}(E_\mathrm{F})$, and (f) experimentally measured longitudinal conductivity $\sigma_\mathrm{xx}$, plotted as functions of LL filling factor $\nu=n\phi_0/B$ and $B$.
		The red-shaded regions indicate symmetry-broken states, where the LL gap exceeds the broadening factor $\Gamma=1.0\,\mathrm{meV}$.
		The purple- and green-shaded regions denote a restored fourfold degeneracy with the integer or half-integer quantum Hall sequence, respectively.
		The black-dashed lines indicate phase boundaries.}
		\label{fig:Fig3}
	\end{figure*}
	
	Indeed, the power of our platform lies in its ability to not only generate a robust SOC in MLG but also to selectively quench it. 
	We engineered two WSe$_2$ layers to be nearly $30^\circ$ relative to the graphene lattice in device N1 ($\theta_\mathrm{t} \approx 28.9^\circ, \theta_\mathrm{b} \approx 28.2^\circ$). 
	In this specific alignment, the global stacking symmetry suppresses the Rashba contribution, while the alignment-dependent interfacial hybridization quenches the valley-Zeeman term~\cite{peterfalvi2022Quantum}, leading to an effective suppression of the overall proximity SOC and preserving the conventional fourfold degeneracy of pristine MLG [Fig.~\ref{fig:Fig2}(b)].
	Consistently, the FFT spectrum is characterized by a single peak at $f_\nu=1/4$ [Fig.~\ref{fig:Fig2}(d)], signifying the effective suppression of the proximity SOC.

	
	Intriguingly, the evolution of the LLs shows a clear phase boundary between symmetry-broken quantum Hall states and fourfold-degenerate states, which shifts progressively toward higher carrier densities with increasing $B$.
	To gain further insight into this degeneracy transition, we analyze the SdH oscillations by relating the resistance features to the LL filling factor $\nu$.
	As shown in Fig.~\ref{fig:Fig3}(a), LLs in the symmetry-broken regime are consistent with the twofold degeneracy $\nu=2N$, where the index $N$ is an integer.
	Beyond the phase boundary, the fourfold degeneracy is restored.
	However, the corresponding quantization sequence transitions to $\nu=4N$, which is distinct from the half-integer sequence $\nu = 4(N+1/2)$ characteristic of pristine MLG (see Supplemental Material~\uppercase\expandafter{\romannumeral9}~\cite{See}).
	The calculated LL spectra in Figs.~\ref{fig:Fig3}(b) and~\ref{fig:Fig3}(c) accurately capture this magnetic-field-induced transition, showing how the initially separated LLs evolve and reorder.
	
	\begin{figure*}[htbp]
		\centering
		\includegraphics[scale=1]{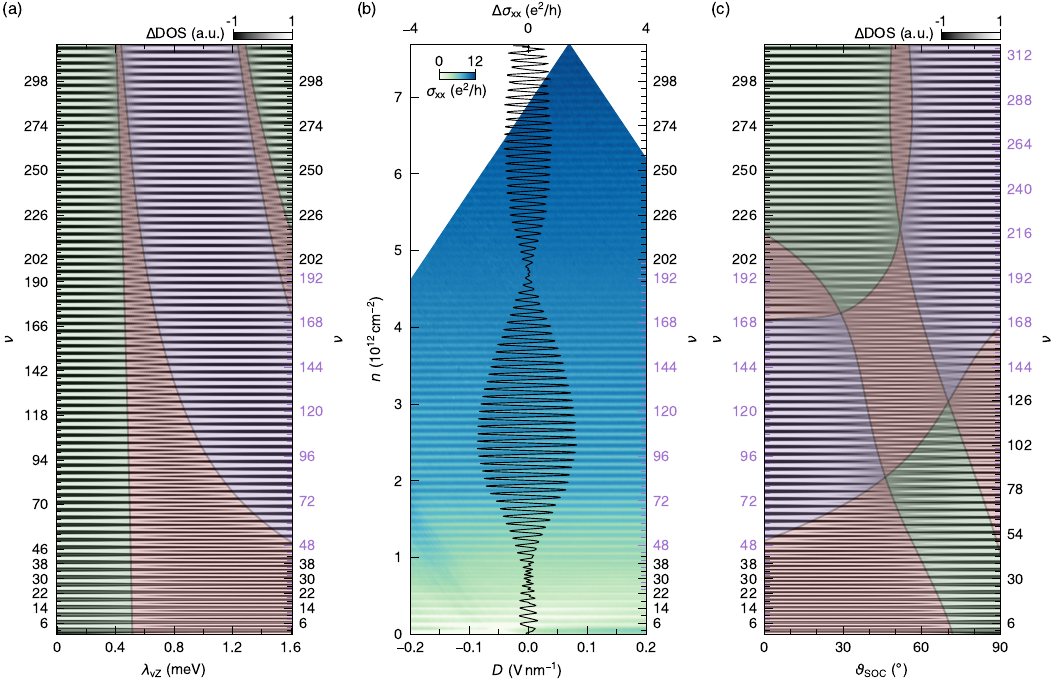}
		\caption{(a) Calculated $\Delta\mathrm{DOS}(E_\mathrm{F})$ as functions of $\lambda_\mathrm{vZ}$ and $\nu$ at $\lambda_\mathrm{R}=0\,\mathrm{meV}$ and $B=1\,\mathrm{T}$.
		(b) Experimentally measured $\sigma_\mathrm{xx}$ as functions of $D$ and $n$ at $B=1\,\mathrm{T}$.
		The line slice of SdH oscillations $\Delta \sigma_\mathrm{xx}$ extracted from $D=0.07\,\mathrm{V\,nm^{-1}}$.
		(c) Calculated $\Delta\mathrm{DOS}(E_\mathrm{F})$ as functions of $\vartheta_\mathrm{SOC} = \arctan(\lambda_\mathrm{R}/\lambda_\mathrm{vZ})$ and $\nu$ at $\lambda_\mathrm{SOC} = 1.6 \,\mathrm{meV}$ and $B=1\,\mathrm{T}$.
		$\Delta\mathrm{DOS}(E_\mathrm{F})$ and $\Delta \sigma_\mathrm{xx}$ are obtained by subtracting the corresponding smooth background.
		The black and purple ticks on the y-axis denote that the quantization sequence is half-integer and integer, respectively.
		The shaded regions in (a) and (c) use the same color scheme as in Fig.~\ref{fig:Fig3}.}
		\label{fig:Fig4}
	\end{figure*}
	
	As schematically depicted in Fig.~\ref{fig:Fig3}(d), this phase transition originates from the evolution of LLs as $B$ increases, driven by a competition between the valley-Zeeman splitting $\Delta_\mathrm{vZ}$ and cyclotron energy spacing $\Delta_{n}$~\cite{castroneto2009Electronic}.
	The conventional Zeeman energy $2E_\mathrm{Z}$ is negligibly small in our measurement ($\sim 0.1\,\mathrm{meV}$ at $1\,\mathrm{T}$) (see Supplemental Material~\uppercase\expandafter{\romannumeral2}~\cite{See}).
	In the case of pure valley–Zeeman SOC, the pristine graphene Dirac cone splits into two pairs of spin-valley-locked sub-bands.
	Under a magnetic field, each sub-band generates its own ladder of LLs separated by the fixed $\Delta_\mathrm{vZ}$~\cite{frank2020Landau}.
	As $B$ increases, the cyclotron energy spacing within each ladder grows as $\Delta_n \propto \sqrt{B}$, while $\Delta_{\mathrm{vZ}}$ remains constant. 
	Consequently, higher-index LLs from the lower-energy sub-band can cross lower-index levels from the upper sub-band, leading to a series of symmetry-enforced LL crossings and reordering~\cite{wang2019Landau, hu2024Topological}.
	
	To benefit visualization of this phase transition, we plot the calculated density of states (DOS) and experimental magnetoresistance as functions of $\nu$ and $B$, as shown in Figs.~\ref{fig:Fig3}(e) and \ref{fig:Fig3}(f), respectively.
	The boundary separates distinct quantization sequences and is well reproduced by the tight-binding calculation.
	Furthermore, this LL reordering provides a sensitive probe of the underlying SOC Hamiltonian. 
	By systematically mapping the phase boundary between the symmetry-broken and degeneracy-restored regimes, we can quantitatively extract the strength of individual SOC components through a combination of tight-binding simulations and high-resolution experimental measurements.
	
	We first isolate the effect of pure valley-Zeeman SOC.
	Figure~\ref{fig:Fig4}(a) presents a calculated phase diagram at $B=1\,\mathrm{T}$, in which $\lambda_\mathrm{vZ}$ is varied from zero to $1.6\,\mathrm{meV}$ with $\lambda_\mathrm{R}=0$.
	The red-shaded region marks the symmetry-broken regime, whose boundary shifts toward lower carrier densities as $\lambda_\mathrm{vZ}$ increases.
	This trend indicates the competition between the valley-Zeeman splitting and the cyclotron energy.
	The increasing $\lambda_\mathrm{vZ}$ requires a larger $\Delta_n$ for LL crossings to restore degeneracy, thereby shifting the boundary toward lower carrier densities at fixed magnetic field~\cite{island2019Spin}.
	This theoretical simulation finds clear experimental validation in Fig.~\ref{fig:Fig4}(b), which displays the measured longitudinal conductivity $\sigma_\mathrm{xx}$ for device D1 as functions of carrier density $n$ and displacement field $D$.
	The observed phase boundaries in Fig.~\ref{fig:Fig4}(b) agree well with the vertical cut through the calculated phase diagram in Fig.~\ref{fig:Fig4}(a) at $\lambda_\mathrm{vZ}=1.6\,\mathrm{meV}$.
	In addition, these boundaries are independent of the applied $D$, confirming that the phase transition is an intrinsic property.

	To further disentangle the overall SOC strength from its composition, we fixed $\lambda_\mathrm{SOC} = 1.6 \,\mathrm{meV}$ and continuously tuned $\vartheta_\mathrm{SOC} = \arctan(\lambda_\mathrm{R}/\lambda_\mathrm{vZ})$ from $0^\circ$ to $90^\circ$, corresponding to pure valley-Zeeman and pure Rashba SOC, respectively.
	The resulting evolution of the phase diagram, shown in Fig.~\ref{fig:Fig4}(c), reveals the distinct symmetry properties of each SOC component.
	In the Rashba-dominated regime, the robust symmetry-breaking region near $\nu=0$ observed in the pure valley-Zeeman case is completely suppressed~\cite{rao2023Ballistic}. 
	Instead, symmetry-broken pockets emerge only within finite ranges of carrier density, which highlights the distinct symmetries of the two SOC terms~\cite{gmitra2016Trivial}.
	The valley-Zeeman component provides a rigid energy splitting of the entire band, whereas the Rashba term modifies the LL spectrum in an energy-dependent manner by coupling spin and momentum.
	
	In this study, the exceptional device quality is further underscored by the clear resolution of LLs at filling factors beyond $\nu=300$, enabling a detailed analysis of the SOC-induced LL evolution.
	The robustness and reproducibility of these findings are further corroborated by a comprehensive analysis of a second device D2, which exhibits consistent valley-Zeeman SOC domination, reproducible LL reordering, and robust TMF peak splitting (see Supplemental Material~\uppercase\expandafter{\romannumeral10}~\cite{See}).
	The consistency between our transport measurements in high-quality vdW heterostructures and tight-binding simulations provides experimental evidence for broader control of the proximity SOC in graphene, establishing a robust platform to design exotic topological and spintronic phases.

		\section*{Acknowledgments}
	The authors would like to thank the International Joint Lab of 2D Materials at Nanjing University for the support, and Yuanchen Co, Ltd (http://www.monosciences.com) for the High-Quality 2D Material Transfer System. 
	G.Y. acknowledges the financial support from the National Key R\&D Program of China (Nos. 2024YFB3715400, 2022YFA1204700, and 2021YFA1400400), the National Natural Science Foundation of China (No. 11974169), the Natural Science Foundation of Jiangsu Province (No. BK20233001), and the support from Nanjing University International Collaboration Initiative. 
	L.W. acknowledges the National Key Projects for Research and Development of China (Nos. 2022YFA1204700 and 2021YFA1400400), Natural Science Foundation of Jiangsu Province (Nos. BK20220066 and BK20233001). 
	R.D. acknowledges the grant from the National Natural Science Foundation of China (No. 12004173). 
	
	\section*{Author contributions}
	G.Y. and R.D. conceived the study; 
	Y.H. prepared the samples with the support from R.D., J.H., D.Z., W.X. and S.W.; 
	J.X. and S.J. performed the transport measurements under the instruction of R.D., F.L. and A.S.M.; 
	S.J. worked out the theoretical model and implemented the numerical simulation with the support from J.J., H.Y. and K.C.; 
	Y.D., Y.L. and X.X. conducted the SHG measurements; 
	K.W. and T.T. grew the hBN crystals; 
	Y.H., S.J. and J.X. analyzed the experimental data with the help of R.D., G.Y., A.S.M. and L.W.; 
	Y.H. and S.J. wrote the manuscript with input from A.S.M., G.Y. and L.W.
	All authors contributed to the discussions and commented on the manuscript.
	
	\section*{Competing interests}
	The authors declare no competing interests.

\end{document}